\documentclass[]{aastex631}

\usepackage{soul}
\usepackage{graphicx}
\usepackage{hyperref}

\submitjournal{ApJL}

\begin{document}

\title{131 and 304 Å Emission Variability Increases Hours Prior to Solar Flare Onset}






\author[0000-0003-3740-9240]{Kara L. Kniezewski}
\affiliation{Air Force Institute of Technology \\
2950 Hobson Way \\
Wright-Patterson AFB, OH 45433, USA}

\author[0000-0002-8767-7182]{E. I. Mason}
\affiliation{Predictive Science Inc. \\
9990 Mesa Rim Rd, Suite 170 \\
San Diego, CA 92121, USA}

\author[0000-0002-5871-6605]{Vadim M. Uritsky}
\affiliation{Catholic University of America \\
620 Michigan Ave., N.E. \\
Washington, DC 20064, USA}
\affiliation{NASA Goddard Space Flight Center \\ 8800 Greenbelt Ave.\\Greenbelt, MD 20771, USA}

\author[0000-0002-9359-6776]{Seth H. Garland}
\affiliation{Air Force Institute of Technology \\
2950 Hobson Way \\
Wright-Patterson AFB, OH 45433, USA}

\begin{abstract}

Thermal changes in coronal loops are well-studied, both in quiescent active regions and in flaring scenarios. However, relatively little attention has been paid to loop emission in the hours before the onset of a solar flare; here, we present the findings of a study of over 50 off-limb flares of GOES class C5.0 and above. We investigated the integrated emission variability for Solar Dynamics Observatory Atmospheric Imaging Assembly channels 131, 171, 193, and 304 Ångstroms for 6 hours before each flare, and compared these quantities to the same time range and channels above active regions without proximal flaring. We find significantly increased emission variability in the 2-3 hours before flare onset, particularly for the 131 and 304 channels. This finding suggests a potential new flare prediction methodology. The emission trends between the channels are not consistently well-correlated, suggesting a somewhat chaotic thermal environment within the coronal portion of the loops that disturbs the commonly-observed heating and cooling cycles of quiescent active region loops. We present our approach, the resulting statistics, and discuss the implications for heating sources in these pre-flaring active regions.

\end{abstract}

\keywords{Active solar corona (1988); Solar flares (1496); Solar physics (1476); Solar coronal loops (1485); Solar corona (1483); Solar coronal mass ejections (310))}

\section{Introduction} \label{sec:intro}

A solar flare is an intense brightening in electromagnetic radiation that may be observed over a wide range of wavelengths such as X-rays (XUV; 0.001 - 10 nm), extreme ultraviolet (EUV; 10 - 120 nm), ultraviolet (UV; 100 - 400 nm), and visible light (380 - 760 nm) \citep[e.g.,][]{Toriumi2019,Tobiska2004}. Flares are often sources of related energetic events in the heliosphere, such as solar energetic particles (SEPs), various spectra of radiation (specifically, XUV/EUV emission), and coronal mass ejections (CMEs). These energetic events, in particular, can have moderate to severe consequences on ground-to-space communication, and our fragile technological infrastructure highlights the importance of early-warning mechanisms \citep{Siscoe2000}.

It is generally accepted that the primary mechanism for energy release in solar flares is magnetic reconnection in the low corona. The Standard Model, which is outlined in \cite{Holman2016}, is the widely accepted empirical view of solar flare progression. Magnetic reconnection occurs over the neutral line, the region dividing magnetic field lines of opposite polarity. Large- or small-scale driving, such as differential rotation or localized energization, cause the field loops to shear. These highly-sheared loops elongate and eventually become unstable, causing the loops to reconnect with one another and building a magnetic flux rope (MFR) above the reconnection site. If the MFR itself becomes unstable, then it can escape the corona as a coronal mass ejection (CME). CMEs, however, are not only caused by solar flares, but can also be related to several other phenomena, such as helmet streamer blowouts. There are also various theories for the early eruptive phase mechanisms. These fall into two major types: ideal instabilities \citep[i.e., torus and kink instabilities;][]{Kliem2006,Torok2004}, and the breakout model of \cite{Antiochos1999}. The main differentiating points between these two models are when the MFR is considered to have formed, and the required magnetic complexity of the structure prior to the eruption. While both of these mechanisms likely play a role in eruptions on the Sun, flare triggers -- the first step that begins destabilization and sets the stage for flare reconnection -- are still an area of active research in which conclusive evidence remains elusive.

As the only available full vector magnetic field data product, many solar flare prediction works have focused on the photospheric magnetic field. Many of these efforts have analyzed the temporal variation of field parameters \cite[e.g.,][]{Leka2003, Fontenla1995, Chen2019, Wang2015} and correlations between field parameters other flare properties \cite[e.g.,][]{Welsch2009, Jing2006, Cui2007}. In recent research of solar flare forecasting, coronal magnetic field has been modeled to investigate the evolution of the 3D magnetic field prior to an eruption. Specifically, nonlinear force-free field (NLFFF) models, although carrying a list of assumptions \cite[see][for a review]{Wiegelmann2007}, have been a popular choice for flare prediction studies to model the magnetically-driven corona due to their simplicity and relatively short computation timescales compared to more sophisticated models \cite[e.g.,][]{Gupta2021, Garland2024, Muhamad2018, Jarolim2023, Yurchyshyn2022}. Forecasting methods incorporating machine learning techniques have also been employed \cite[e.g.,][]{Bobra2015, Huang2018, Florios2018, Nishizuka2018}. Ultimately, all forecasting efforts currently face challenges to create a successful prediction, including but not limited to data availability and the rarity of strong flare events \citep{Camporeale2019}. 

EUV emission above flare-active active regions are the focus of the work discussed here. EUV data products have only recently started to be analyzed for flare forecasting purposes. \cite{Dissauer2023} and \cite{Leka2023} examine SDO AIA channels for a range of plasma and magnetic parameters prior to flares for a large sample of active regions with the extensive AIA Active Region Patches database. It has been reported that coronal loops expand and active region upflows increase in the day leading up to a solar flare \citep{Imada2014}. Case studies which examine coronal EUV emission have previously found small scale brightening \cite[e.g.,][]{Sterling2001, Joshi2021} and coronal dimming \citep[e.g,][]{Lee2013,Harra2013,Kerr2021} prior to an eruptive event. Few of these case studies examine enough examples to differentiate between idiosyncratic and representative behavior, however, which is necessary to both understand and predict such events on a larger scale.

In this paper, we study the emission from off-limb loops above active regions in the hours leading up to significant flares (which we define as those of at least GOES class C5.0), to focus on the thermal changes in the corona prior to flare onset. Coronal loops, the major magnetic structures of the X-ray corona, have been observed and modeled for decades, dating back to the Skylab era \cite[e.g.,][]{Tousey1973}. Loops are anchored in the photosphere and have length scales on the order of Mm. In the magnetically driven ($\beta \ll$ 1) and high temperature ($\sim 10^6$ K) corona environment, the completely ionized gas interacts with the coronal field. While we cannot measure coronal magnetic fields directly, we can infer it through the plasma it confines \citep[e.g.,][and many others]{Moore2001,Antiochos1998,Su2011,Temmer2010}. As the fundamental structures of the corona, coronal loops have been a focus of many efforts to search for the source of coronal heating \cite[e.g.,][]{Aschwanden2000, Klimchuck2006, Ascwanden2007, Spadaro2003}. 

In a non-flaring context, coronal loops can be assumed to be quasi-static and can be modeled hydrostatically. \citet{Rosner1978} modeled static coronal loops, and developed
scaling laws that relate the peak temperature at the loop apex to the uniform heating rate, constant pressure, and loop length. Observations of coronal loops in hotter
active regions do agree well with this quasi-static picture; observations show that emission of these loops varies very little over the timescale of hours, which is exceedingly long in comparison with the timescales of conduction and radiative loss \citep{Antiochos2003}. However, heat fluctuations resulting from impulsive heating due to nanoflares \citep{Parker1988, Klimchuck2006} or the steady but spatially constrained heating characterizing thermal non-equilibrium \citep{Antiochos1991} can result in thoroughly dynamic loop emissions. Numerous numerical simulations and models have investigated the myriad possible coronal heating mechanisms characterized by different temporal and/or spatial profiles of the heating function \citep{Reale2019, Klimchuk2008, Lionello2013}.

The temporal variation in emission from these quiescent coronal loops prior to solar flares is the focal point of this study. The longer loops above an active region are often found to be in thermal nonequilibrium, which has well-observed cycles of a few hours in length of slow loop heating and localized catastrophic cooling, often resulting in condensations known as coronal rain \citep{Auchere2018,Froment2020,Sahin2023}. While performing analysis for a previous paper, we noticed that these cycles often seemed to undergo perturbations in the time before flare onset that resulted in a distinct increase or decrease in loop brightness, warranting further investigation. Solar eruption triggers and their timescales are still poorly understood. A number of models have been proposed for individual observational aspects of field destabilization and eruption initiation \cite[see][for an extensive review]{Forbes2006} and agree well with individual portions of an eruption, but the majority lack a numerical threshold for initiation. For this purpose, we examine the coronal loop emissions for six hours prior to the flare. Common quiescent loop signatures such as thermal nonequilibrium generally occur on 2-3 hour time scales, and loop cooling times are on the scale of one hour; 6 hours was chosen as a window that would show several thermal cycles if present, as well as any shorter-term changes to them due to the upcoming flare. In this paper, we integrate coronal loop emission within a flaring active region located on the limb. Changes in loop emission can provide substantial information regarding the evolution of the coronal magnetic field and heating mechanisms prior to a flare. We propose that the greatly enhanced variation of these coronal loops can be utilized to predict a significant flare event hours prior to the impulsive phase. 

A detailed description of our data selection for pre-flare and control active region loops is described in Section \ref{data}. Our loop integration and statistical analysis methods are described in Section \ref{methods}. Section \ref{sec:results} discusses the results and statistics of the emission integration trends. We provide theory and implications behind our emission variation analysis in Section \ref{conclusions}, including a discussion for potential avenues for future work in solar flare forecasting applications. Ultimately, the success in flare prediction efforts lie in uncovering what mechanisms trigger a flare and their timescales. We discuss how coronal loop emission variation can potentially supply information for flare early-warning.

\section{Data Selection and Methods} \label{data}
\subsection{Data Selection}

This study uses solar flare data compiled in the X-ray Telescope (XRT) Flare Catalog (\url{https://xrt.cfa.harvard.edu/flare_catalog/}) and the Heliophysics Event Knowledgebase (HEK; \url{https://www.lmsal.com/hek/index.html}). The XRT Flare Catalog contains basic information from all flares during the Hinode \citep{Kosugi2007} mission from December 19th, 2006, through July 31st, 2023, and the HEK flare catalog was queried for events detected by the Geostationary Operational Environmental Satellite (GOES). Both catalogs were queried for events between 2011 January 1 and 2022 December 31, since the Solar Dynamics Observatory (SDO) began operation part way through 2010. Only flares greater than C5.0 were considered, in accordance with the findings of \cite{Mason2022} that there was a significant energy difference between flares above and below the C5.0 threshold. Events were therefore also discounted if a flare of magnitude C5.0 or greater occurred in the same active region within 6 hours of the event of interest. Finally, only flares on the limb ($\pm$ 850 arcsec) were considered, so that coronal loop observations were as clear and unobstructed as possible. After these filters were applied, we selected all X-class flares and a pool of M-class flares that were spread throughout magnitude and throughout Solar Cycle 24 and the rise of Solar Cycle 25. Events were examined when they were selected, and if they needed to be eliminated for any reason (e.g, data dropouts, other coronal structures contaminating the line of sight, etc.), a new flare was chosen from the XRT list. The final event selection includes 4 X-class flares, 25 M-class flares, and 24 C-class flares. The full data set used in this study is available at \href{https://zenodo.org/doi/10.5281/zenodo.13696124/}{zenodo.org/doi/10.5281/zenodo.13696124/}, and includes the flare start time, GOES class, host active region, eruptivity, presence or absence of a 131 Å spike, and event location in heliographic coordinates. Most of this information is also reproduced in Table \ref{tab:events}, below.

Event selection for the non-flaring ``control'' active region cases was similar. We looked for active regions which were of similar magnetic complexity at the time they were observed on the limb (the active regions in our flaring cases were predominantly of the beta-gamma and beta-gamma-delta Hale classes, as is common with flare-active active regions), and which were sufficiently isolated in both latitude and longitude to not contaminate the line of sight. We used the magnetic complexity classification from the nearest on-disk magnetogram observations; if the active region had been observed on the eastern limb, we ensured that there were not significant signatures of emergence observed in EUV emission between the limb observation and the magnetogram. An additional criterion for our flare-quiet cases was to select active regions which did have recorded flares, but that had at least a 12-hr span with no flares of C or above. This allowed us to select a 6-hr window with at least a 3-hr buffer before and after it, capturing the quietest possible conditions for our measurements. While this significantly narrowed our available observations, these stringent requirements ensured that our results were not skewed by inherent differences between the types of active regions, but zeroed in on differences in the loops due to the upcoming flaring activity. We selected 30 such non-flaring time windows in 30 unique active regions, and the data for those is also available at the Zenodo DOI listed above.

\begin{table}[ht]
\centering
\resizebox{0.6\textwidth}{!}{%
\begin{tabular}{||c|c|c|c|c||}
\hline
\textbf{Flare Onset} & \textbf{GOES Class} & \textbf{NOAA AR} & \textbf{CME} & \textbf{131 Spike} \\ \hline
2011-02-24T07:23:00 & M3.5 & 11163 & Yes & No  \\ \hline
2011-03-08T18:08:00 & M4.4 & 11165 & Yes & No  \\ \hline
2011-12-14T13:18:00 & C5.8 & 11367 & No  & No  \\ \hline
2011-12-28T14:17:00 & C7.2 & 11389 & No  & Yes \\ \hline
2012-03-02T17:29:00 & M3.3 & 11429 & Yes & No  \\ \hline
2012-03-09T19:59:00 & C9.7 & 11432 & Yes & No  \\ \hline
2012-03-23T16:31:00 & C6.5 & 11445 & No  & No  \\ \hline
2012-05-05T09:30:00 & C6.8 & 11476 & No  & Yes \\ \hline
2012-05-17T01:25:00 & M5.1 & 11476 & Yes & No  \\ \hline
2012-07-09T08:05:00 & C7.9 & 11515 & No  & Yes \\ \hline
2012-07-19T04:17:00 & M7.7 & 11520 & Yes & Yes \\ \hline
2012-08-06T00:39:00 & C9.4 & 11542 & Yes & No  \\ \hline
2012-08-18T00:24:00 & M5.5 & 11548 & Yes & No  \\ \hline
2012-10-08T11:05:00 & M2.3 & 11589 & No  & Yes \\ \hline
2012-10-20T18:05:00 & M9.0 & 11598 & Yes & Yes \\ \hline
2012-11-30T16:56:00 & C5.4 & 11620 & No  & No  \\ \hline
2013-03-20T00:39:00 & C5.5 & 11698 & No  & No  \\ \hline
2013-04-26T06:20:00 & C7.0 & 11726 & No  & Yes \\ \hline
2013-04-26T22:09:00 & C5.7 & 11726 & No  & Yes \\ \hline
2013-05-03T17:24:00 & M5.7 & 11739 & Yes & No  \\ \hline
2013-07-02T17:45:00 & C7.1 & 11785 & Yes & No  \\ \hline
2013-07-03T07:00:00 & M1.5 & 11787 & Yes & No  \\ \hline
2013-07-29T23:07:00 & C6.3 & 11800 & No  & No  \\ \hline
2013-10-26T19:24:00 & M3.1 & 11884 & Yes & No  \\ \hline
2013-12-19T23:06:00 & M3.5 & 11934 & Yes & No  \\ \hline
2014-01-07T22:16:00 & C7.4 & 11947 & No  & No  \\ \hline
2014-04-25T00:17:00 & X1.3 & 12035 & Yes & No  \\ \hline
2014-05-05T18:10:00 & C8.0 & 12056 & No  & Yes \\ \hline
2014-05-06T04:19:00 & C7.1 & 12051 & Yes & No  \\ \hline
2014-06-15T07:50:00 & C7.0 & 12085 & Yes & Yes \\ \hline
2014-08-24T12:00:00 & M5.9 & 12151 & Yes & Yes \\ \hline
2014-10-02T18:49:00 & M7.3 & 12173 & Yes & No  \\ \hline
2014-11-03T22:15:00 & M6.5 & 12205 & Yes & No  \\ \hline
2014-11-04T07:59:00 & M2.6 & 12205 & Yes & Yes \\ \hline
2014-11-14T07:42:00 & C5.4 & 12209 & No  & No  \\ \hline
2014-12-10T17:07:00 & C5.9 & 12222 & Yes & No  \\ \hline
2015-01-21T11:32:00 & C9.9 & 12268 & Yes & No  \\ \hline
2015-05-05T22:05:00 & X2.7 & 12339 & Yes & No  \\ \hline
2015-08-30T02:01:00 & M1.4 & 12403 & No  & Yes \\ \hline
2015-09-28T20:15:00 & C7.6 & 12423 & No  & No  \\ \hline
2016-07-24T06:09:00 & M2.0 & 12567 & No  & Yes \\ \hline
2017-08-20T19:20:00 & C9.4 & 12672 & Yes & No  \\ \hline
2017-09-10T15:35:00 & X8.2 & 12673 & Yes & No  \\ \hline
2020-11-29T12:34:00 & M4.4 & 12790 & Yes & No  \\ \hline
2021-04-22T20:02:00 & C8.5 & 12817 & No  & Yes \\ \hline
2021-05-07T18:43:00 & M3.9 & 12822 & Yes & No  \\ \hline
2021-07-03T14:18:00 & X1.5 & 12838 & Yes & No  \\ \hline
2022-04-17T02:00:00 & M1.8 & 12994 & No  & No  \\ \hline
2022-05-11T18:13:00 & M2.6 & 13004 & Yes & No  \\ \hline
2022-09-16T09:44:00 & M7.9 & 13098 & No  & Yes \\ \hline
2022-09-17T20:32:00 & M2.6 & 13098 & No  & No  \\ \hline
2022-10-14T09:20:00 & M1.3 & 13112 & Yes & No  \\ \hline
2022-12-31T02:27:00 & C7.9 & 13180 & No  & No  \\ \hline
\end{tabular}%
}
\caption{Limb flares which were used in this study. This information, with additional location information, is available online at \href{https://zenodo.org/doi/10.5281/zenodo.13696124/}{zenodo.org/doi/10.5281/zenodo.13696124/}. Where active region numbers were missing from the XRT flare catalog (generally due to an active region erupting prior to receiving its official NOAA designation), we have manually identified the region.}
\label{tab:events}
\end{table}

\subsection{Observational Methods}
\begin{figure}[ht]
    \centering
    \includegraphics[scale=0.68]{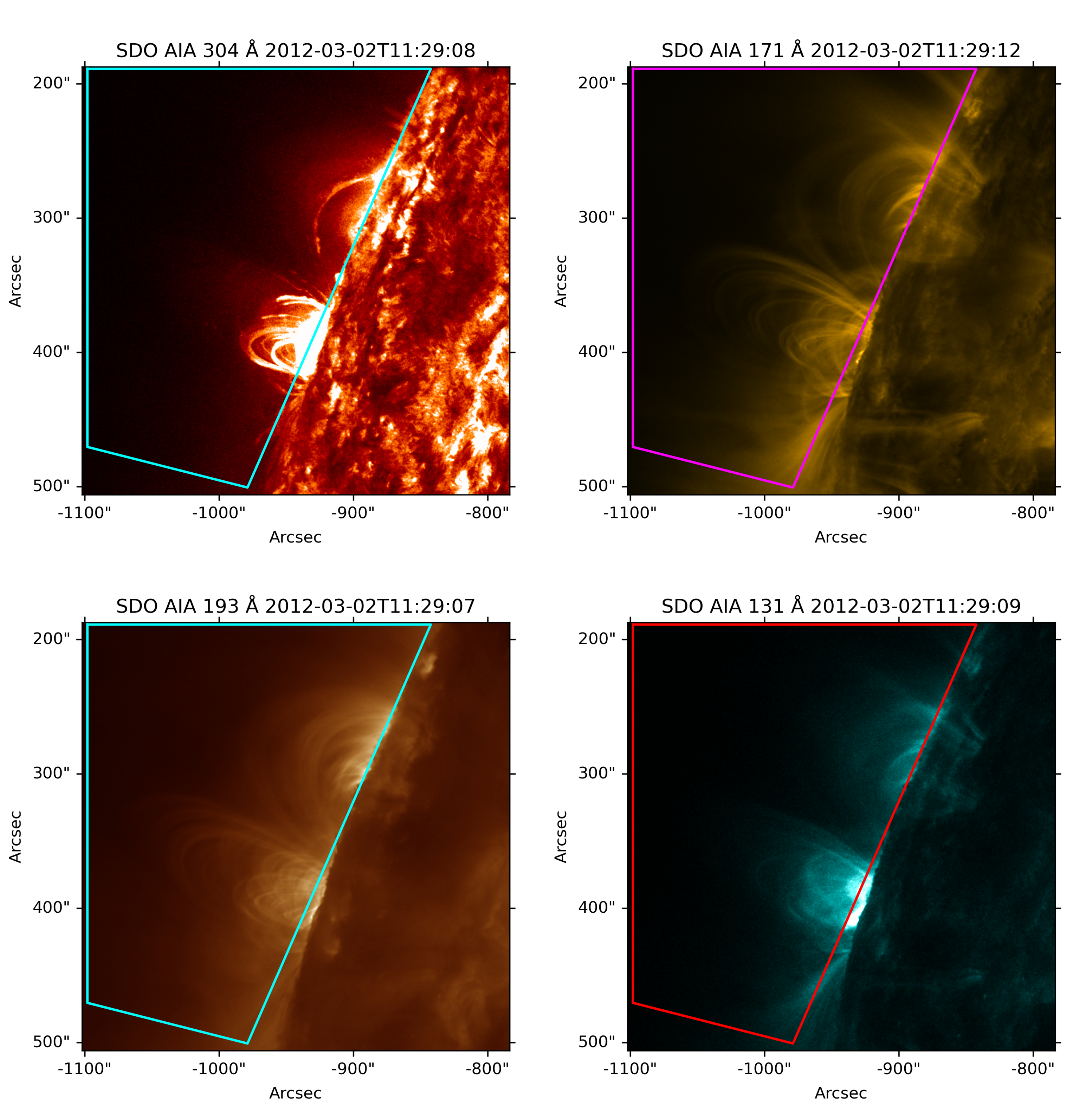}
    \caption{Example emission boxes for a M3.3 flare in 304 Å (top left), 171 Å (top right), 193 Å (bottom left), and 131 Å (bottom right). The boxes represent the emission integration region, which shares the same dimensions across each of the wavelength channels and remained constant throughout the six hour integration period.}
    \label{fig:aia4channels}
\end{figure}

The pre-flare coronal loop emissions were analyzed with SDO Atmospheric Imaging Assembly (AIA) data for 6 hr prior to a solar flare, at 3 min cadence. Coronal loops emit strongly in the X-ray and EUV, appearing as bright arcs in these wavelength channels with their footpoints anchored in the photosphere. These arch-like structures are easily identifiable on the limb and were the central features used in this study. Since our interest is in the aggregate thermal trends in the longer loops of active regions prior to flares, it was unnecessary for us to attempt to isolate individual loops. Instead, we selected trapezoidal regions which captured the greatest off-limb region possible for each active region while minimizing emission from any structures in front of or behind the active region along the line of sight. To investigate the range of emission over the six-hour period, we used 131 Å, which peaks at two temperatures (400,000 K and 10 MK plasmas) and is often used to study flaring regions within the corona; 171 Å, which is particularly used to examine coronal loops structures and absorption/emission of coronal rain condensations around 1 MK; 193 Å, for its highlights of the hot corona and flaring plasmas at 1.25 MK; and 304 Å, which is typically used to examine cool plasma at 50,000 K \citep{Lemen2012}. An example structure of loops in each of these wavelengths is presented in Figure \ref{fig:aia4channels}, where the well-defined loops can be clearly seen in 171 Å and cool condensations persist within the loops in 304 Å. Note that 131 was chosen specifically for its bimodal response function: it is often easy to visually differentiate the sources of emission in 131 by whether the emission is ``fuzzy" or diffuse (the hot contribution) or more compact and localized (the cool contribution). Since we are not using it for a calculation such as a differential emission measure, trends in this channel can be manually investigated, and thus help interpret similar spikes in the other channels. In fact, we found that a significant number of cases had sharp increases in 131 emission well before the other channels; this is discussed in more detail in Section \ref{conclusions}.

We also determined whether each flare was eruptive using the same methods as \cite{Mason2022}. The HEK was queried one hour following the start time of each flare to determine if a CME was reported within a reasonable time after the flare and if its location was related to the flare. To confirm if a CME candidate was from a flare event of interest, CMEs were visually identified in the SOHO/LASCO CME Catalog\footnote{This CME catalog is generated and maintained at the CDAW Data Center by NASA and The Catholic University of America in cooperation with the Naval Research Laboratory. SOHO is a project of international cooperation between ESA and NASA.} (\url{https://cdaw.gsfc.nasa.gov/CME_list/}).  

\section{Methodology}\label{methods}

We processed all of the SDO AIA data to level 1.5 using the standard tools from the {\tt SunPy} Python package, version 5.1.3 (\url{https://doi.org/10.5281/zenodo.11180141}) of the SunPy open source software package \cite{Barnes2020}. Since the active regions which we study here are on the limb and we were only interested in short-term observations of the off-limb loops, de-rotation was not required in the processing pipeline. We chose not to apply any background-subtraction or other image processing techniques to the data because we wanted to work with the image data in a form close to that from the near-real-time pipeline, to test for future applicability as an operational tool.

We manually selected polygonal regions, such as the one shown in Figure \ref{fig:aia4channels}. The guidelines we used in selecting the polygons were to maximize the off-limb loops within the polygon while minimizing contamination from other nearby active regions to the north and south of the target region. As discussed previously, during the event selection process we excluded active regions which had other active regions in close proximity to the east or west, such that they contaminated the line-of-sight integration.

We used the polygonal regions to construct a mask for each wavelength, and integrated the emission within the mask at each time, normalized to the exposure time. These integrations were saved to JSON files for easy retrieval for plotting and further calculations. The same process was applied to the non-flaring control cases.

\begin{figure}[ht]
    \centering
    \includegraphics[width=\linewidth]{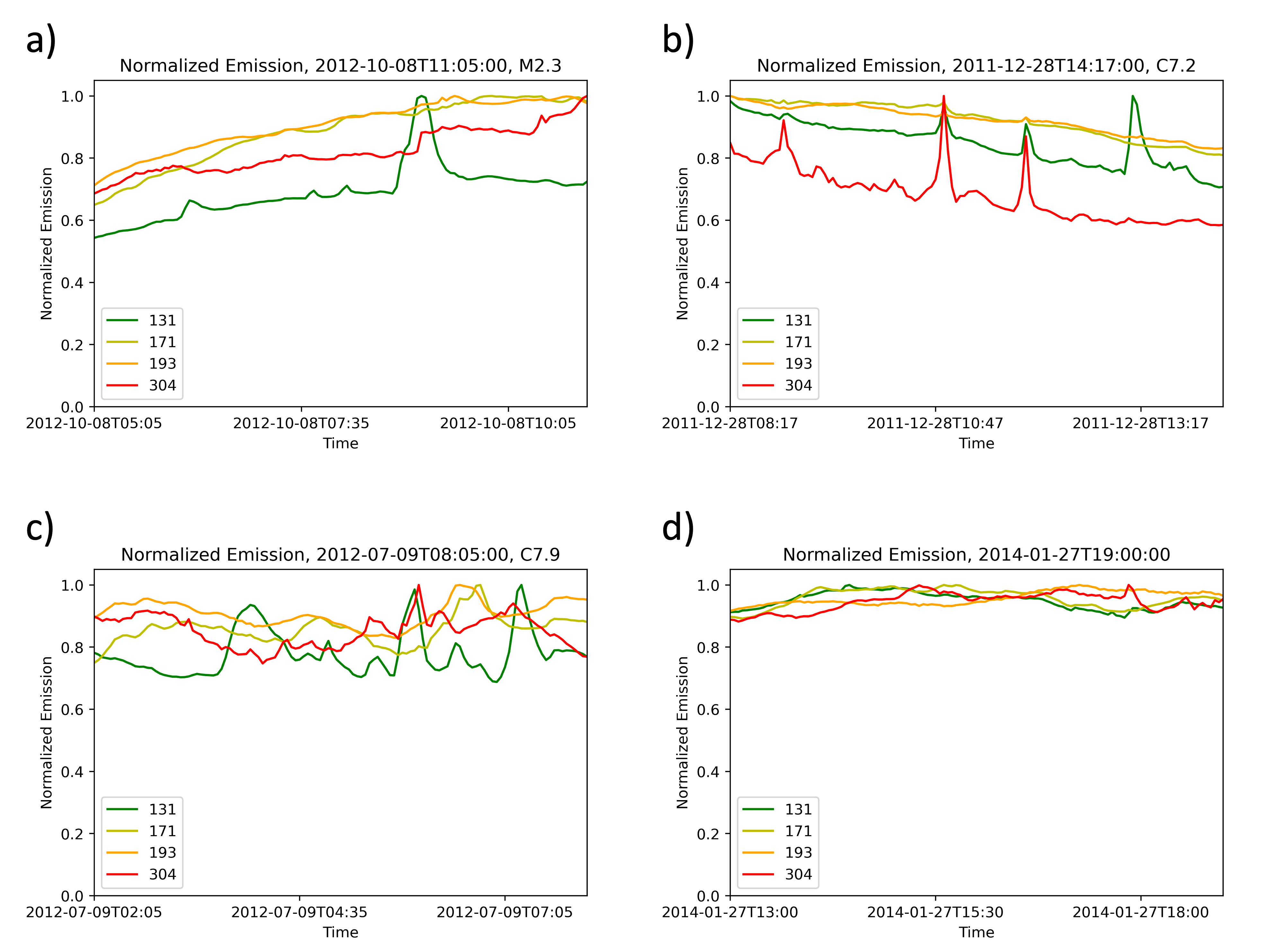}
    \caption{a) Example normalized integrated emission plot for the 6 hr prior to a M2.3 flare on 2012-10-08T11:05:00 showing a general increase in all channels; of particular note is the significant spike in the 131 emission several hours before the flare onset, which is not seen in the other 3 wavelengths. b) an analogous plot for a C-class flare, showing a general decrease in all channels. c) an analogous plot for a different C-class flare, showing poor correlation between the channels. d) an analogous plot for a non-flaring time period, showing significantly less variation in the emission levels during a 6-hr period.}
    \label{fig:flare_nonflare_comp}
\end{figure}

We began by plotting all four wavelengths for each case, with the emission for each normalized to 1. Figure \ref{fig:flare_nonflare_comp} shows sample integrated emission curves of all 4 channels studied for 4 cases: three flaring cases, each with different overall trends, and one non-flaring case. It is important to note that these trends are not all of equal frequency; those flaring cases with either a slow, uniform increase \textit{or} decrease in emission in all 4 channels account for less than 30\% of the flaring cases. The rest have more complex emission patterns, and a very few are relatively steady. All of the integrated emission plots are available at the Zenodo link. Then, we began the statistical analysis by computing detrended fluctuation analysis and structure functions. We also compute the standard deviation of the total emission in each channel, and then derive the ratio of the standard deviations between flaring and non-flaring active regions. The purpose of the standard deviation ratios was to compare overall emission variability between flaring and non-flaring active regions. We conducted these statistics for time windows beginning 1 hr before the flare and increasing up to the full 6-hr window, to determine what time before the flares the active regions showed the best indication of coming activity and to rule out the possibility that flare-active active regions simply have an inherent tendency towards greater emission variability.

These analyses addressed both trends within individual cases, as well as considering each flare class in comparison to non-flaring cases. We would like to emphasize that there were not enough X-class flares that met our criteria from which to draw meaningful statistics (there were 4 total). We present our results for the X-class cases for completeness and out of the significant interest in the most energetic flares, but \textit{we caution the reader against drawing specific conclusions} from the statistics shown below.

\section{Results}\label{sec:results}

\subsection{Statistical Analysis}

\begin{figure}
    \centering
    \includegraphics[width=\linewidth]{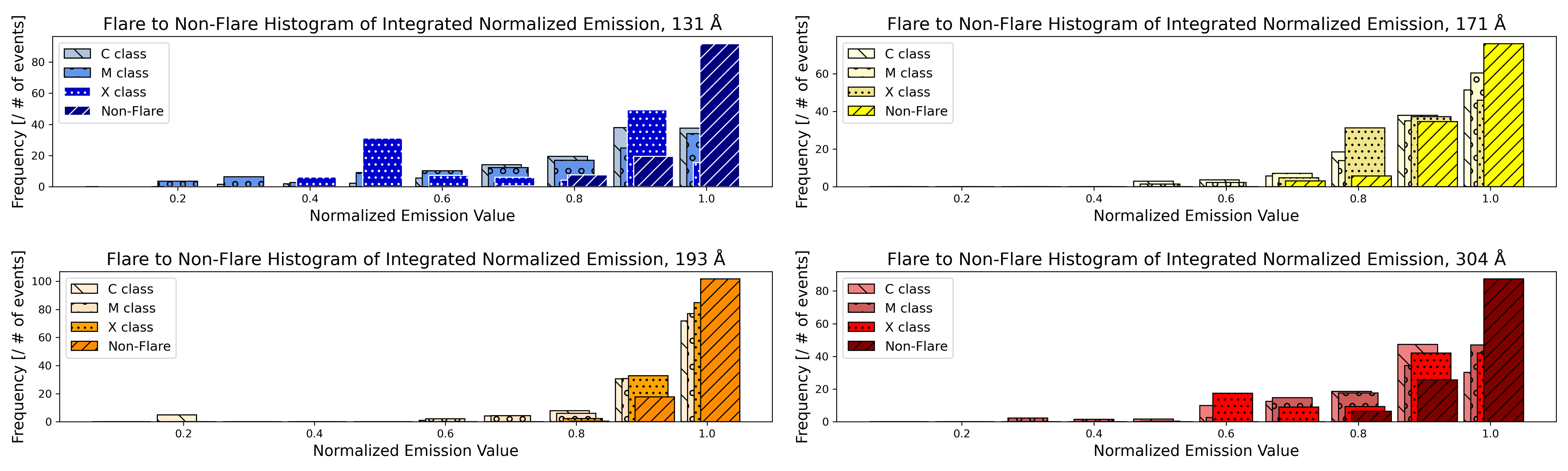}
    \caption{Stacked histogram plots, one for each SDO AIA channel investigated. The bins cover the full available range of values in the normalized integrated emission plots, and the different shades and patterns denote C flares, M flares, X flares, and non-flaring cases.}
    \label{fig:hists}
\end{figure}

To assess the differences between the flaring and non-flaring cases and determine their significance, we took a combination of approaches. The most visual and intuitive method is shown in Figure \ref{fig:hists}, where the normalized channel frequencies are plotted as histograms for each GOES class and channel against the non-flaring cases. Each set of cases (the three flare classes and the non-flaring cases) have been normalized to the number of cases in that pool, so that the columns are comparable. It is clear from these histograms that the the distribution of values is much greater for the flaring cases than it is for the non-flaring cases in 131 and 304 Å. The same trend is evident but markedly weaker in 171 and 193 Å. This shows that the variability is significantly greater for 131 and 304 Å in the hours leading up to a flare, though this particular measure does not help pinpoint the most relevant time range which can indicate a flare is likely.

\begin{figure}[htb!]
    \centering
    \includegraphics[width=\linewidth]{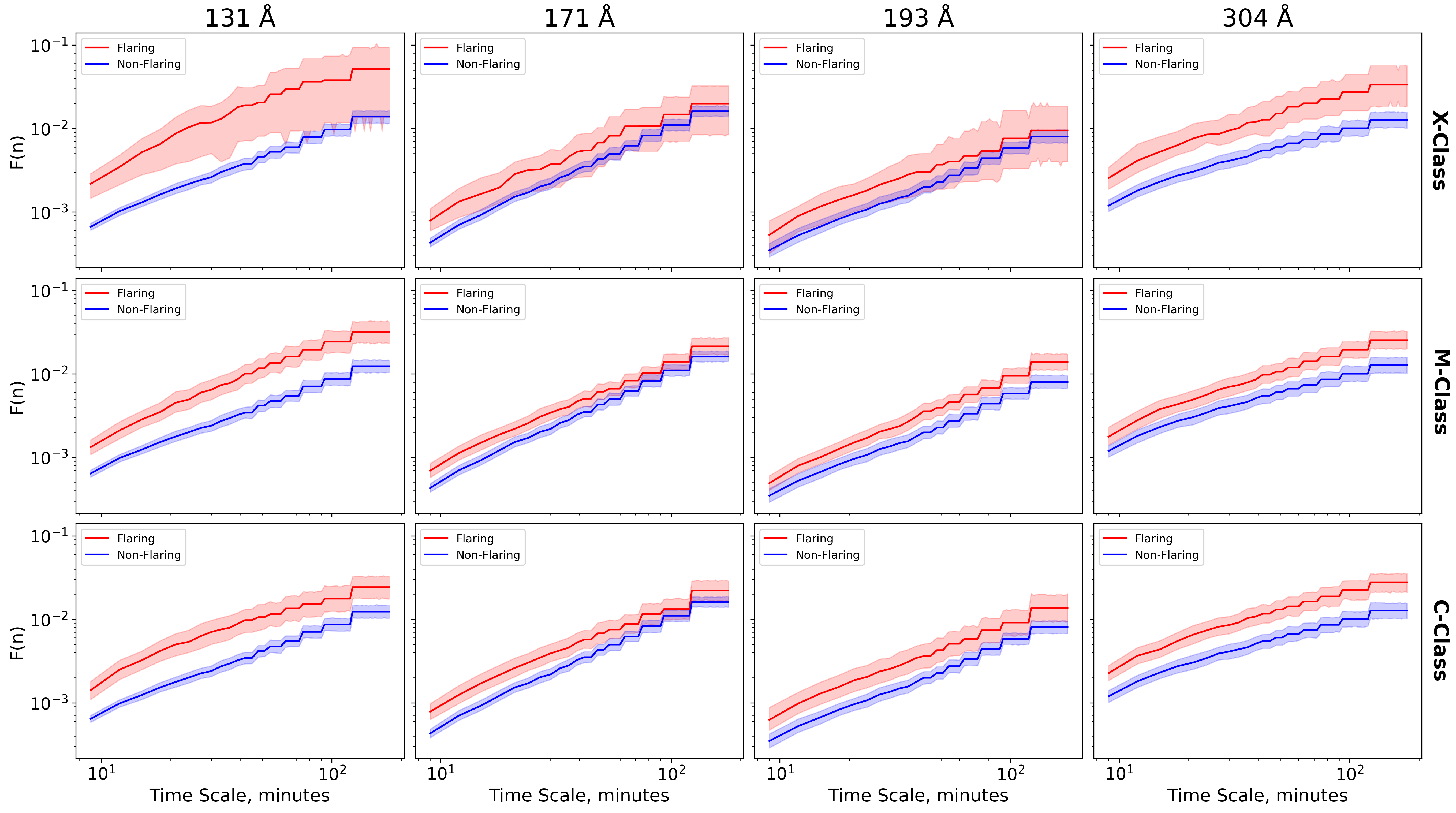}
    \caption{Detrended fluctuation analysis of the flaring (red) and non-flaring (blue) time series for each AIA channel (columns) and GOES flare class (rows). Note that computations were done for normalized data in each channel.}
    \label{fig:DFA}
\end{figure}

\begin{figure}[htb!]
    \centering
    \includegraphics[width=\linewidth]{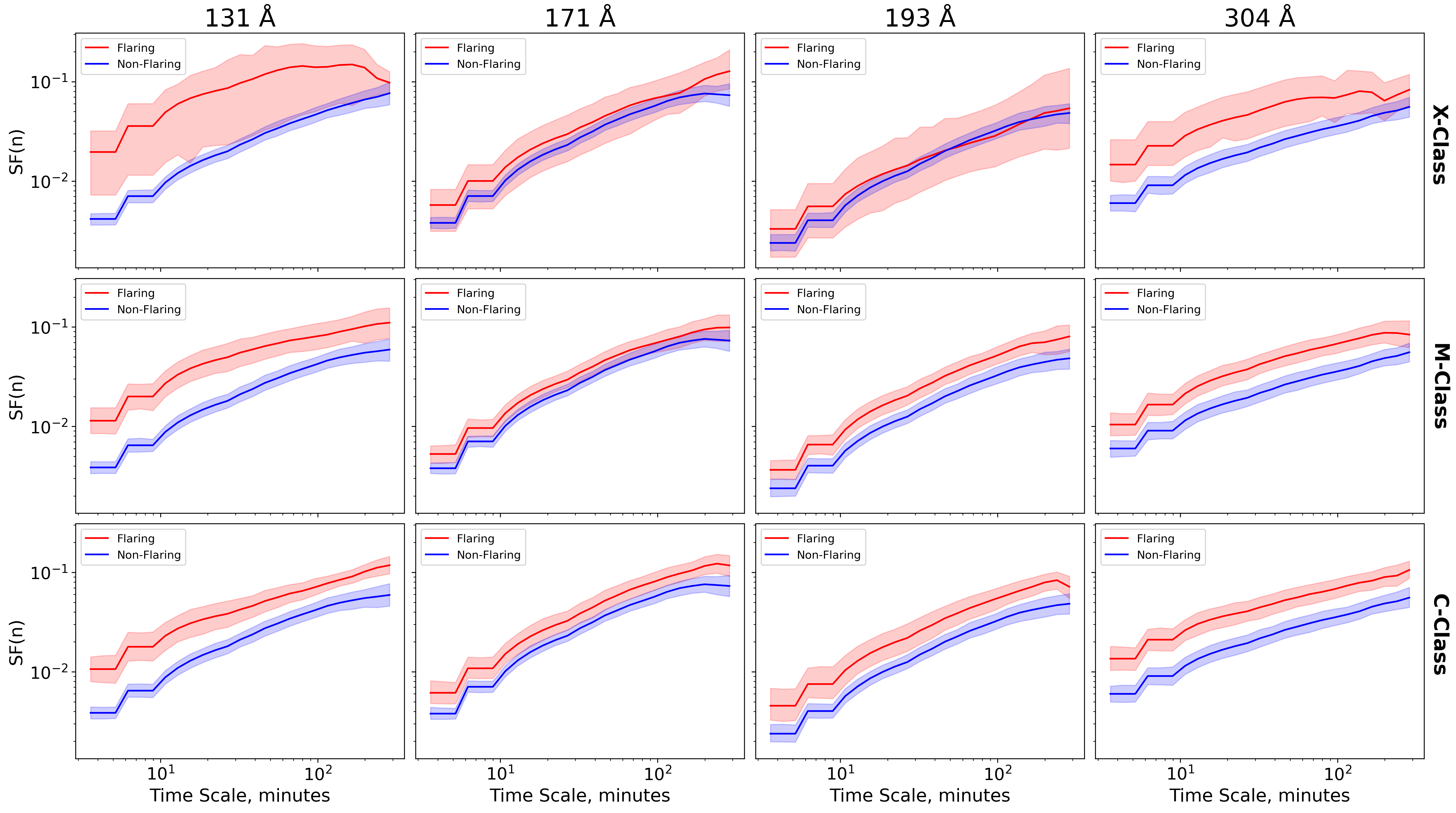}
    \caption{Structure function analysis of the flaring (red) and non-flaring (blue) time series for each AIA channel (columns) and GOES flare class (rows). Note that computations were done for normalized data in each channel.}
    \label{fig:SF}
\end{figure}

Additionally, we applied structure functions \cite[SF;][]{Rutman1978} and detrended fluctuation analysis \citep[DFA;][]{Peng1994} to the integrated emission curves to quantify the differences between the various flaring cases and the non-flaring cases. The fluctuation function from DFA methods is defined as
\begin{equation}
    F(n) = \sqrt{< \left( y\left(t_{i}\right) - Y\left(t_{i}\right) \right)^2 >}
\end{equation}

\noindent
where the brackets $<...>$, denote averaging, $y\left(t_{i}\right)$ is the time series signal (the integrated emission curves for each class and channel, in our analysis) which is divided into linear time window bins, $n$, and $Y\left(t_{i}\right)$ is a linear regression fit to $y\left(t_{i}\right)$ within a given bin. The second-order structure function is defined as

\begin{equation}
    SF(n) = \sqrt{< \left(|y\left(t_{i}\right) - y\left(t_{j}\right) \right|)^2 >}
\end{equation}

\noindent
with $n = t_{i} - t_{j}$. We chose to compute the scales as exponentially increasing bins in the SF fluctuation calculations; this provides uniform binning on a log-log scale and is better at handling sparse tails in distributions. The DFA and SF fluctuation functions for the normalized emission data are plotted in Figures \ref{fig:DFA} and \ref{fig:SF}, divided by channel and GOES class. It is important to note that the slopes of the SF plots are consistently lower for the flaring cases than for the quiet cases in 131 and 304 Å; this shows that short-term fluctuations contribute more significantly to the overall variability than long-term shifts. This hints at more dynamic thermal processes, rather than limb observation effects or inherent large-scale differences in the active regions themselves. 

Across all of the wavelength channels, the flaring characteristic variability is larger than the non-flaring variability, and it gradually increases with an increasing time scale. Specifically, 131 and 304 Å show the most statistically significant differences between the flaring and non-flaring fluctuations, and do not have overlapping confidence intervals. While the fluctuation functions for pre-flare 171 and 193 Å are greater than the non-flaring fluctuations, any statistical significance in their variability is significantly weaker than 131 and 304 Å in the time prior to a flare. The spacing between the flaring and non-flaring DFA and SF variability curves also seems to scale with GOES flare class consistently for 131 Å. Furthermore, the difference between pre-flare and non-flare fluctuations appears to increase with increasing flare strength. This trend requires corroboration from a larger sample of pre-flare and non-flare emissions, particularly for the X-class cases.

\subsection{Standard Deviation Ratios}

\begin{figure}[htb!]
    \centering
    \includegraphics[width=\linewidth]{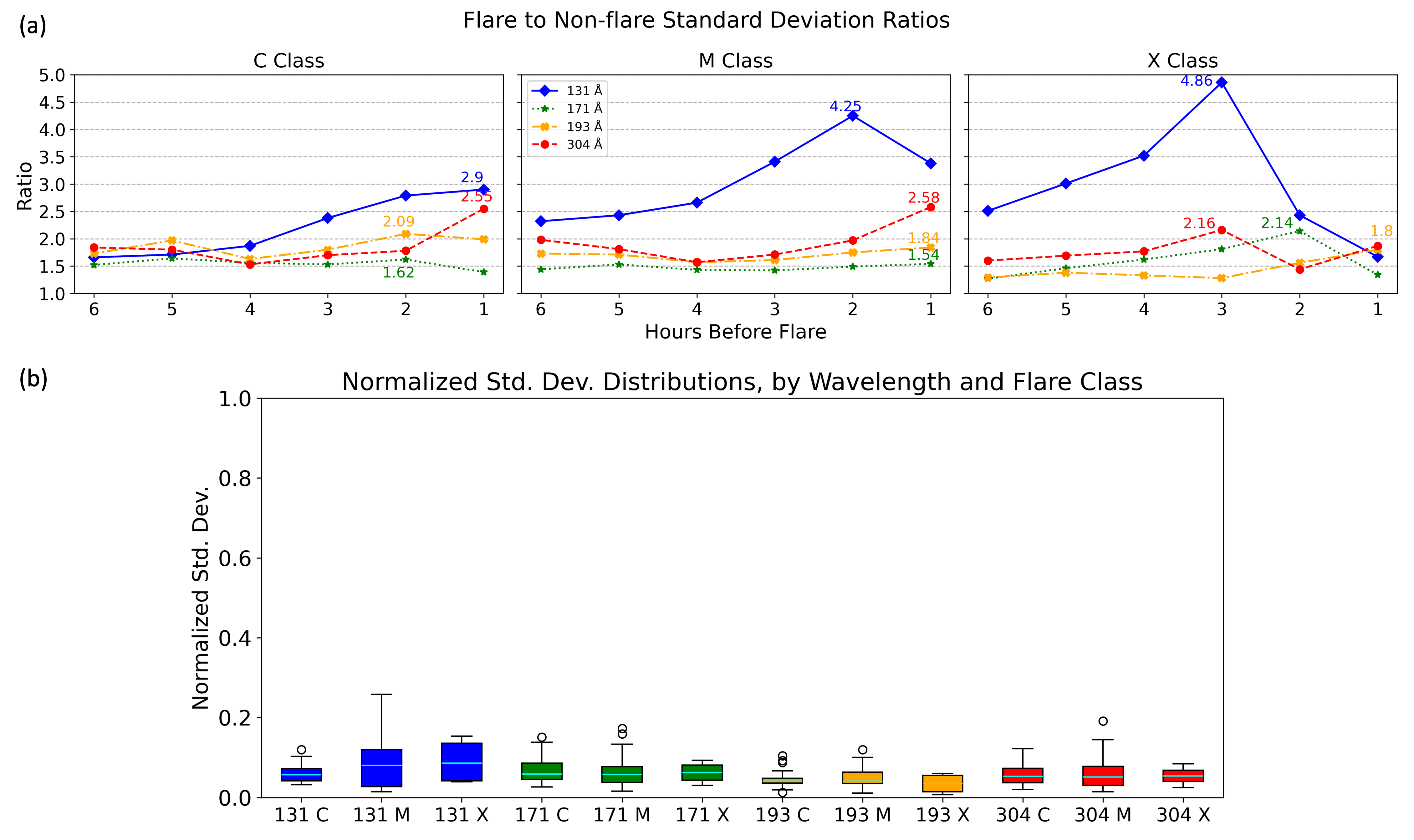}
    \caption{(a) Cumulative hourly standard deviation ratios by GOES flare class and cumulative number of hours before flare onset. The y-axis for all three plots is 1 to 5, and the 0-point on the x-axis corresponds to the timestamp of flare onset. The maximum value is labeled with its respective value for each channel. (b) box and whisker plot showing the distribution of normalized standard deviations, by GOES class and AIA channel. The median value for each dataset is represented by a cyan line, and outliers are shown as black circles.}
    \label{fig:stddevs}
\end{figure}

In order to address the time period before a flare in which the fluctuations were most apparent, we calculated the ratio of the standard deviations of each flare class to the standard deviations of the non-flaring cases. The population standard deviation is defined as

\begin{equation}
    \sigma = \sqrt{\frac{\sum_{i} |x_{i} - \bar{x}|^2}{N}}
\end{equation}

\noindent
where $x_{i}$ is an individual value of the emission, $\bar{x}$ is the emission average, and $N$ is the total number of observations. For equal sample sizes, the pooled standard deviation can be calculated by

\begin{equation}
    \sigma_{p} = \sqrt{\frac{\sum_{i}\sigma_{i}^2}{N_{p}}}
\end{equation}

\noindent
where $\sigma_{i}$ is an individual population's standard deviation and $N_{p}$ is the number of populations in the pool. A pooled standard deviation is useful for combining the standard deviations of a group (e.g, flare class) into one common standard deviation. Therefore, the ratio between the flaring and non-flaring pooled standard deviations can simply be calculated as

\begin{equation}
    R = \frac{\sigma_{p}[flare]}{\sigma_{p}[flare-quiet]}
\end{equation}

\noindent
with $\sigma_{p}[flare]$ and $\sigma_{p}[flare-quiet]$ signifying the flaring and non-flaring pooled standard deviations. These ratios were calculated for the full number of cumulative hours before the flare (i.e., for one hour before the flare, for the two hours before the flare, etc. with each larger time period including the data from the shorter one before it). The results are presented in Table \ref{tab:stddevs} and plotted in Figure \ref{fig:stddevs}a. The bottom panel of this figure shows the normalized standard deviations for each GOES class and AIA channel; the small spread and few outliers show that the standard deviation ratios are not dominated by one or two anomalously volatile events in each pool, but are rather representative of an overall trend across flaring cases.

\begin{table}[htb!]
    \centering
    \begin{tabular}{||c|c|c|c|c||}
    \hline
        \multicolumn{5}{||c||}{\textbf{1 HOUR}} \\ \hline
        ~ & 131 & 171 & 193 & 304  \\ \hline
        \textit{C-Class} & 2.90 & 1.39 & 1.99 & 2.55  \\ \hline
        \textit{M-Class} & 3.38 & 1.54 & 1.84 & 2.58  \\ \hline
        \textit{X-Class} & 1.67 & 1.34 & 1.80 & 1.87  \\ \hline
        \multicolumn{5}{||c||}{} \\ \hline
        \multicolumn{5}{||c||}{\textbf{2 HOUR}} \\ \hline
        ~ & 131 & 171 & 193 & 304  \\ \hline
        \textit{C-Class} & 2.79 & 1.62 & 2.09 & 1.78  \\ \hline
        \textit{M-Class} & 4.25 & 1.49 & 1.75 & 1.97  \\ \hline
        \textit{X-Class} & 2.43 & 2.14 & 1.56 & 1.44  \\ \hline
        \multicolumn{5}{||c||}{} \\ \hline
        \multicolumn{5}{||c||}{\textbf{3 HOUR}} \\ \hline
        ~ & 131 & 171 & 193 & 304  \\ \hline
        \textit{C-Class} & 2.38 & 1.53 & 1.80 & 1.70  \\ \hline
        \textit{M-Class} & 3.41 & 1.42 & 1.61 & 1.71  \\ \hline
        \textit{X-Class} & 4.86 & 1.81 & 1.28 & 2.16  \\ \hline
        \multicolumn{5}{||c||}{} \\ \hline
        \multicolumn{5}{||c||}{\textbf{4 HOUR}}\\ \hline
        ~ & 131 & 171 & 193 & 304  \\ \hline
        \textit{C-Class} & 1.87 & 1.56 & 1.63 & 1.53  \\ \hline
        \textit{M-Class} & 2.66 & 1.43 & 1.57 & 1.57  \\ \hline
        \textit{X-Class} & 3.52 & 1.62 & 1.33 & 1.77  \\ \hline
        \multicolumn{5}{||c||}{} \\ \hline
        \multicolumn{5}{||c||}{\textbf{5 HOUR}}\\ \hline
        ~ & 131 & 171 & 193 & 304  \\ \hline
        \textit{C-Class} & 1.71 & 1.64 & 1.97 & 1.80  \\ \hline
        \textit{M-Class} & 2.43 & 1.53 & 1.71 & 1.81  \\ \hline
        \textit{X-Class} & 3.01 & 1.46 & 1.38 & 1.69  \\ \hline
        \multicolumn{5}{||c||}{} \\ \hline
        \multicolumn{5}{||c||}{\textbf{6 HOUR}}\\ \hline
        ~ & 131 & 171 & 193 & 304  \\ \hline
        \textit{C-Class} & 1.66 & 1.52 & 1.73 & 1.84  \\ \hline
        \textit{M-Class} & 2.32 & 1.44 & 1.73 & 1.98  \\ \hline
        \textit{X-Class} & 2.51 & 1.27 & 1.29 & 1.60 \\ \hline
    \end{tabular}
    \caption{Table showing the values of the standard deviation ratios of each GOES flare class studied here to the non-flaring cases, by the number of hours of data included before flare onset.}
    \label{tab:stddevs}
\end{table}

The 131 and 304 Å channels consistently have the highest peak ratios, and their peaks all exceed a factor of 2 over non-flaring cases. Across all wavelengths studied in the C and M classes, there is a near-uniform steady increase in variability from 4 hr before flare onset, peaking in the 1-2 hr before onset (the sole exception to this trend is 171 Å for C-class flares, which dips near flare onset). It is worth noting that the ratio for flare cases never drops to 1 or below, indicating that flare-active active regions may have some level of variability that is more or less constantly above that of a flare-quiet active region. Nevertheless, there is a marked positive slope to the 131 and 304 Å ratios from 4 hr before the flare onset. The X-class cases, while a statistically insignificant pool, show an even more marked increase for 131 Å but a decrease in the ratios for all wavelengths leading up to flare onset (with the exception of 193 Å).

\begin{figure}
    \centering
    \includegraphics[width=\linewidth]{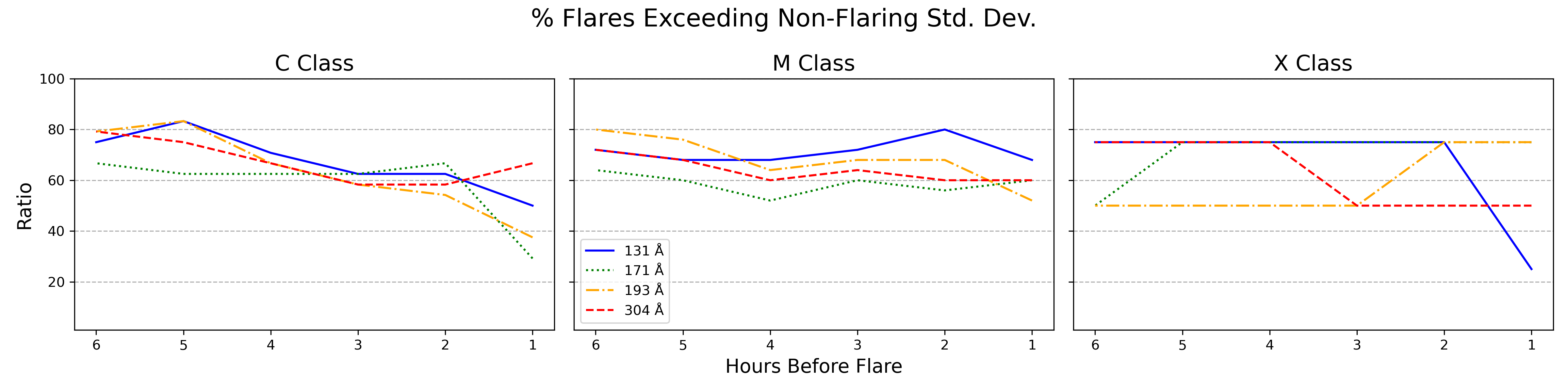}
    \caption{This figure graphs the percentage of flares whose integrated emission standard deviations exceed those of the non-flaring cases for the cumulative 6-hr time range studied here. The exceptionally small event pool for X-class flares causes the poor data quality in the graph on the far right, and is included only for completeness.}
    \label{fig:percs}
\end{figure}

Figure \ref{fig:percs} shows the percentage of flares in each class whose integrated, normalized emission standard deviations exceed those of the non-flaring cases, over the same cumulative time range. For C- and M-class flares, the standard deviations in all four channels exceed the quiet cases in the range of 60-80\% of the time during the 6 hr time frame. In C-class flares, the 131 and 193 Å data are the best predictors around 5 hr pre-flare, while 171 peaks at 2 hr and 304 at 6 hr. In M-class flares, 131 Å peaks around 2 hr pre-flare while the other three peak at 6 hr. While these statistics are promising, they should be combined with the results from Figure \ref{fig:stddevs} for the most actionable result. For M-class flares in particular, this shows that the 131 Å data is the best predictor 2 hr pre-flare, with an 80\% accuracy and a standard deviation ratio of over 4.

\subsection{131 Å Spikes}

\begin{figure}
    \centering
    \includegraphics[width=\linewidth]{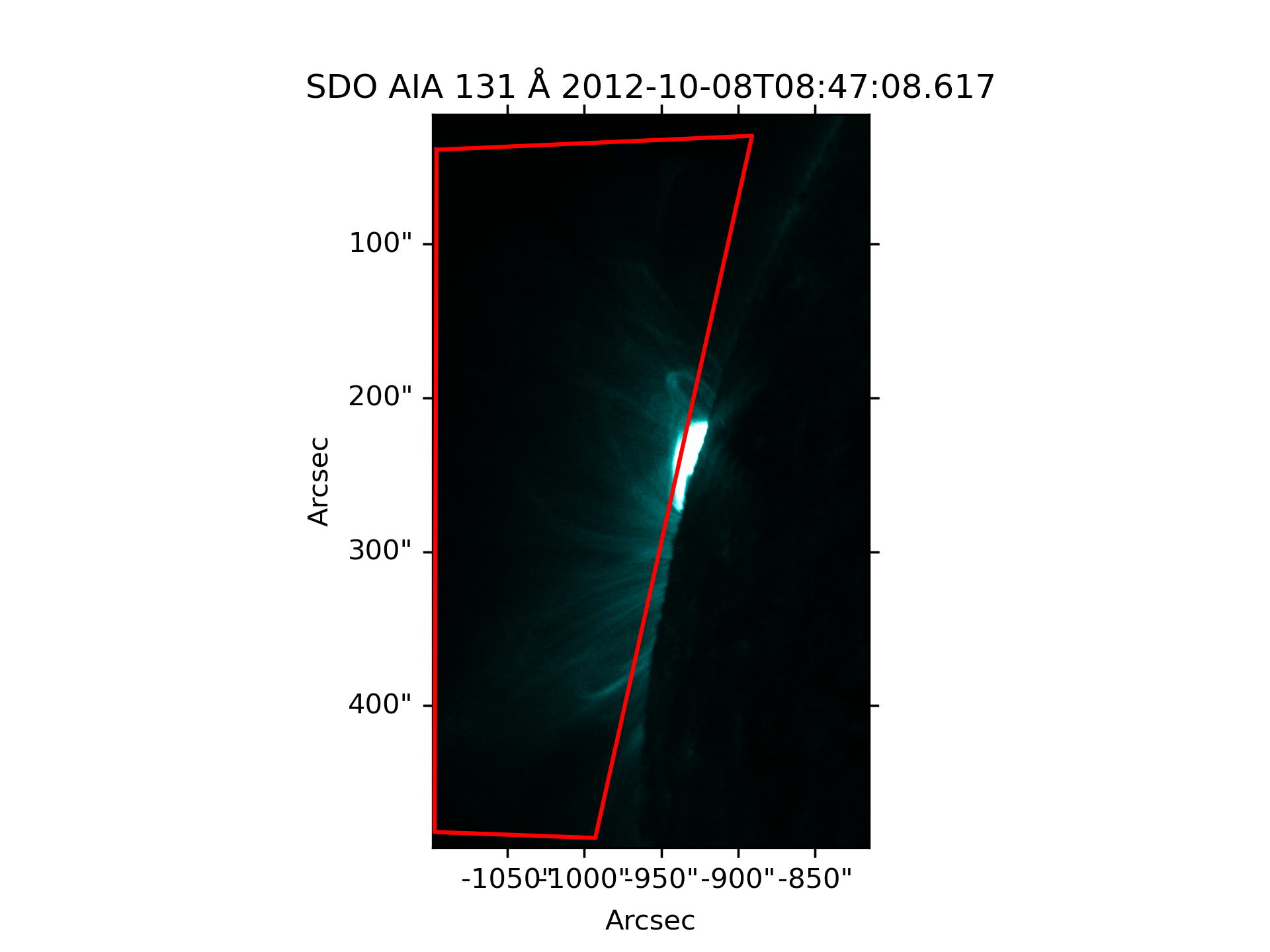}
    \caption{A still frame showing the emission enhancement that caused a spike in the respective 131 Å integration, with the integration polygon included for clarity. The animation online shows the full 6-hr span of the integrations in a 5 s movie, spanning the times from 2012-10-08T05:05:08 to 2012-10-08T11:02:08.}
    \label{fig:131_spikes}
\end{figure}

While plotting the normalized integration for the flaring cases, it became apparent that a non-negligible number of cases had distinct peaks superimposed on the overall trend in the 131 Å channel data. We generated movies with our selected polygons overlaid to pinpoint any regions which were being captured in the integrations. One such example can be seen in Figure \ref{fig:131_spikes}, which is also an animation online. 

This figure is a representative case; most of the time these emission spikes originate from very short, compact loop groups near the center of the active region, and the greatest emission often comes from the loop apexes. Since we selected the polygons to exclude the limb, it is apparent that not all of the emission is even captured by the polygon, but there was enough to cause a distinct enhancement. It is therefore also probable that some 131 spikes were missed due to our polygon selection criteria. Some, but not all, of the spikes are associated with weaker flares (under C5.0). However, out of 16 cases with clear 131 Å enhancements, 11 of them occurred before confined flares. The breakdown for these enhancements by GOES class is summarized in Table \ref{tab:131_spikes}. For both C- and M-class flares, the 131 Å spike is over three times as likely to herald a confined flare as an eruptive one. We discuss possible causes and implications for this in the following section.

\begin{table}[h]
\center
\begin{tabular}{||c|c|c||}
\hline
\textit{} & CME      & No CME   \\ \hline
C         & 1 (13\%) & 7 (44\%) \\ \hline
M         & 4 (21\%) & 4 (67\%) \\ \hline
X         & 0 (0\%)  & 0 \\ \hline
\end{tabular}
\caption{Frequencies and percentages of 131 Å spike prevalence for eruptive and non-eruptive flares.}
\label{tab:131_spikes}
\end{table}

\section{Discussion and Conclusions}\label{conclusions}

This is, to our knowledge, the first study of its kind to analyze and compare the overall emission variability of various temperature EUV channels between imminently flaring and non-flaring active regions. Various studies have investigated methods by which magnetic processes of varying time frames prepare a flux rope or overlying active region loops for flare reconnection \citep[][among others]{Antiochos1998,Torok2005,Titov2008,Green2009,Nindos2020,Kliem2021}, many of which rely on magnetic models and force-free extrapolations of magnetograms. This study provides evidence that some process directly affects the thermal conditions above an active region in the hours leading up to a major flare. 

The nature of this process is currently unknown. While some localized loops could become brighter or dimmer due to solar rotation during this time frame (and many likely do both), the optically thin nature of the EUV emission and the selection of as many of the active region's long loops as possible serve to minimize this effect. Normal quiescent coronal loop heating theory would dictate that a loop undergoing a cooling process would see hotter channel emission decrease as cooler channel emission increased (as is commonly seen with thermal nonequilibrium in active region loops), while a loop undergoing heating would see the opposite effect. A loop whose density was suddenly increased (through so-called chromospheric evaporation) or decreased (as in rapid loop growth, sometimes observed before flares) could see the channels react in unison, due to the $n^2$ dependence of most EUV emission; either of these could be possible explanations for the slow increase or decrease in emission seen in Figure \ref{fig:flare_nonflare_comp}a and b. However, uniform gradual increases/decreases in emission occur in less than a third of the total flare cases studied here, and there is no consistency between uniform gradual changes in emission and GOES class. Often even within a single 6-hour window, the trends among the 4 channels can change. This implies that there is likely more than one mechanism at play in controlling active region loop emission leading up to a flare, and that they act on different time scales. 

Even though processes like thermal nonequilibrium occur within individual loops, there is evidence that nearby groups of loops undergo similar heating and cooling cycles in near-synchrony, which should make the integrated emission relatively coherent. Our data, on the contrary, appear to be superpositions of multiple signatures; this suggests that an additional process is heating localized regions in the corona and disturbing the typical thermal patterns. Several future studies are necessary to better understand these results; one to determine any spatial and temporal patterns and another to extend the data to on-disk regions, if possible. A study including on-disk active regions would also allow for the magnetic evolution of the active regions to be included in the analysis; this study was designed to analyze the coronal loops' emission without contamination from the transition region or chromosphere, so no magnetic data was available for these observations.

Our major findings can be summarized as follows:

\begin{itemize}
    \item \textbf{The variability of EUV emission in active region loops before a significant flare (GOES C5.0 or greater) is significantly higher than the variability in a similar active region that is not imminently flaring.} This is particularly true for the 131 and 304 Å channels, but the flare/quiet ratios for 171 and 193 Å are also both above 1 for all times studied. Furthermore, the variability exceeds non-flaring variability in the range of 60-80\% of the time for the C- and M-class flares we studied.
    \item \textbf{This variability increases from 4 hr before flare onset, peaking in the 1-2 hr before onset.} This provides a solid threshold by which to test potentially-flaring regions, and presents greater lead time than current flare prediction schemes. It also appears that the variability peak moves earlier as the GOES class increases, but a larger pool of X-class flares is required to confirm this trend.
    \item \textbf{The enhanced pre-flare variability is observed across all studied time scales and is the most pronounced at the shortest scales (of the order of several minutes), hinting at the impulsive transient nature of the underlying physical process.} The highly localized compact brightenings observed in many of the cases in 131 Å and the general localized nature of 304 Å emission support this finding.
    \item \textbf{We found that a non-negligible proportion of significant flares were preceded by major enhancements in the 131 Å emission, and that these were three times more likely before a confined flare than before an eruptive flare.} It is likely -- due to the guidelines we used in selecting the integration regions -- that there are additional cases of such 131 Å spikes, since these brightenings tended to occur in very short loops. 
\end{itemize}

The normalized integration plots for the flare cases showed a range of patterns, as shown in Figure \ref{fig:flare_nonflare_comp}: in some, there was good coherence between the channels and they all increased or decreased leading up to the flare; in others, the channels showed little to no coherence at all; and in a few, all 4 channels appeared to be mostly stable. Both the first and second scenario resulted in high standard deviation ratios, while the third would result in a smaller ratio, explaining our overall findings. However, there are clearly important physical processes that would generate such varied behavior, and a future study would need to tackle these underlying mechanisms. One possible explanation for the less-coherent responses is that localized currents in the corona allow reconnection high up the loops. This could spark isolated and chaotic heating events than more pervasive processes like wave heating or high-frequency nanoflare heating near loop footpoints. We intend to conduct a subsequent study using dates for both a flaring and non-flaring active region when one of the STEREO spacecraft \citep{Wuelser2004,Kaiser2008} was in quadrature with SDO, in order to include both the magnetic field information for current extrapolations alongside the off-limb EUV channel integrations (using all four EUVI channels, most of which overlap with our current selection, with the notable exception of 131 Å). We will also use these dates to compare the results of an off-limb integration to an on-disk integration.

The spikes in the 131 Å emission due to the compact loop groupings presents an intriguing challenge. The main question is why such a high proportion of these precede confined flares. One potential solution is that there is only strong reconnection occurring in these short arcades that cause the enhancement in 131 Å, which are not long enough to create a coherent MFR. Another is that there is some inherent relationship or positive feedback loop between the amount of energy released into these loops to create such enhanced emission at very high temperatures, and the strength of the overlying (or strapping) field (which is commonly believed to be a central factor in creating confined flares). Extending this study to on-disk flares will allow the magnetic field to be analyzed, and determine the field strength in the region where these bright loops are located.

The simplicity of the region selection and calculation method we developed for this project lends itself well to automation, and the strong results of the standard deviation ratios show that this method is useful for flare prediction. Figures \ref{fig:stddevs} and \ref{fig:percs} show that this method could predict flares 2-6 hr ahead of time with an accuracy of 60-80\%. These statistics are for each channel analyzed independently, without considering more complex inter-dependencies of the various channels upon each other. They may improve if a variability in time offset between channels is identified. Typical active region loop heating cycles have been well-studied using time lag analysis \citep{Viall2012,Viall2016,Viall2017,Barnes2021}, and normal time lag ranges and distributions are known. We plan to apply the same methodology to these off-limb observations. If the distributions or values are significantly different, it would provide evidence for and data on any localized heating occurring in these regions.

In this paper, we have presented promising preliminary results of a new flare analysis method. Our findings indicate that there are energetics occurring many hours before the impulsive phase of a solar flare begins, which drive significant thermal shifts in the loops on the time scale of hours. Given the range of trends across the various cases we studied, it is likely that there is more than one mechanism at work driving these changes. However, this analysis may prove useful for both understanding pre-flare changes and potentially for predicting significant flares.

\begin{acknowledgments}
The authors would like to thank Pete Riley for valuable discussions on the interpretation of the results. The views expressed are those of the authors and do not reflect the official guidance or position of the United States Government, the Department of Defense (DoD) or of the United States Air Force.
\end{acknowledgments}

\bibliography{bibliography}{}
\bibliographystyle{aasjournal}



\end{document}